\documentclass{ieeetmlcn}

\usepackage{cite}
\usepackage{amsmath,amssymb,amsfonts}
\usepackage{graphicx,color}
\usepackage{textcomp}
\usepackage{xcolor}
\usepackage{hyperref}
\usepackage[normalem]{ulem}

\usepackage{amsmath,amsfonts}
\usepackage{algorithmic}
\usepackage{array}
\usepackage{cuted}
\usepackage{amsmath}
\usepackage{soul}

\usepackage{soulutf8}  
\hypersetup{hidelinks=true}
\usepackage[linesnumbered,lined,boxed,commentsnumbered]{algorithm2e}
\usepackage{algorithmic}
\def\BibTeX{{\rm B\kern-.05em{\sc i\kern-.025em b}\kern-.08em
    T\kern-.1667em\lower.7ex\hbox{E}\kern-.125emX}}
\AtBeginDocument{\definecolor{tmlcncolor}{cmyk}{0.93,0.59,0.15,0.02}\definecolor{NavyBlue}{RGB}{0,86,125}}
\definecolor{myyellow}{RGB}{240, 240, 100}  
\definecolor{mylightblue}{RGB}{173, 216, 230}  
\definecolor{lightgreen}{RGB}{144, 238, 144}
\definecolor{lightbrown}{RGB}{181, 101, 29}

\usepackage[numbers,sort&compress]{natbib}

\usepackage{textcomp}
\usepackage{stfloats}
\usepackage[font=footnotesize,labelfont=bf]{caption}
\usepackage{subcaption}
\usepackage{mathtools}

\usepackage{stfloats}
\usepackage{url}
\usepackage{verbatim}
\usepackage{graphicx}

\def\BibTeX{{\rm B\kern-.05em{\sc i\kern-.025em b}\kern-.08em
    T\kern-.1667em\lower.7ex\hbox{E}\kern-.125emX}}
\usepackage{balance}
\usepackage{array,multirow,graphicx}
\newcolumntype{L}{>{\centering\arraybackslash}m{1.8cm}}

\title{Dual Actor DDPG for Airborne STAR-RIS Assisted Communications}
\author{Danish Rizvi and David Boyle}

{\vspace{-2mm}}
\affil{Imperial College London, UK}
{\vspace{-5mm}}

\begin{document}

\begin{abstract}
This study departs from the prevailing assumption of independent Transmission and Reflection Coefficients (TRC) in Airborne Simultaneous Transmit and Reflect Reconfigurable Intelligent Surface (STAR-RIS) research. Instead, we explore a novel multi-user downlink communication system that leverages a UAV-mounted STAR-RIS (Aerial-STAR) incorporating a coupled TRC phase shift model. Our key contributions include the joint optimization of UAV trajectory, active beamforming vectors at the base station, and passive RIS TRCs to enhance communication efficiency, while considering UAV energy constraints. We design the TRC as a combination of discrete and continuous actions, and propose a novel Dual Actor Deep Deterministic Policy Gradient (DA-DDPG) algorithm. The algorithm relies on two separate actor networks for high-dimensional hybrid action space. We also propose a novel harmonic mean index (HFI)-based reward function to ensure communication fairness amongst users. For comprehensive analysis, we study the impact of RIS size on UAV aerodynamics showing that it increases drag and energy demand. Simulation results demonstrate that the proposed DA-DDPG algorithm outperforms conventional DDPG and DQN-based solutions by 24\% and 97\%, respectively, in accumulated reward. Three-dimensional UAV trajectory optimization achieves 28\% higher communication efficiency compared to two-dimensional and altitude optimization. The HFI based reward function provides 41\% lower QoS denial rates as compared to other benchmarks. The mobile Aerial-STAR system shows superior performance over fixed deployed counterparts, with the coupled phase STAR-RIS outperforming dual Transmit/Reflect RIS and conventional RIS setups. These findings highlight the potential of Aerial-STAR systems and the effectiveness of our proposed DA-DDPG approach in optimizing their performance.

\end{abstract}

\maketitle
\RestyleAlgo{ruled}

\section{Introduction}
\IEEEPARstart {T}he pursuit of next-generation advancements beyond 5G (B5G) and sixth-generation (6G) networks is underway, driven by the pressing demand for high spectrum efficiency, energy-efficiency, ultra-low latency, and extensive coverage \cite{alghamdi2020intelligent}. As an offshoot to the requirements, technologies such as Reconfigurable Intelligent Surfaces (RIS) alongside with increasing use of UAVs in wireless networks have emerged as front runners in shaping the future landscape of wireless communications~\cite{alsarayreh2023intelligent}. Reconfigurable intelligent surfaces (RISs) are innovative structures composed of passive elements capable of dynamically altering the phase of electromagnetic waves. These surfaces act as "smart mirrors" that can manipulate the propagation environment of wireless signals. By adjusting the phase of reflected waves, a RIS can enhance signal strength, mitigate interference, and establish effective communication links between transmitters and receivers \cite{kaur2024survey}.

While conventional RIS technology has shown promise in enhancing wireless communication, it is limited by its ability to serve users only in the reflection space~\cite{lu2021aerial}. STAR-RIS overcomes this limitation by enabling simultaneous transmission and reflection, thus providing 360$^\circ$ coverage \cite{xu2021star} and mitigating the RIS orientation problem. STAR-RIS technology represents a paradigm shift in wireless communication systems, offering advantages over conventional RIS including extended service coverage, increased beamforming flexibility, and the ability to provide 360$^\circ$ coverage \cite{9570143}. This transformative potential enhances the attractiveness of adopting STAR-RIS solutions in next-generation wireless networks.

Furthermore, the choice of phase shift model employed in STAR-RIS implementations significantly influences system performance. Using a practical coupled phase shift model \cite{9838767} as opposed to an independent phase shift model degrades performance due to interdependence of transmit and reflect coefficients, but is considered more realistic of the two \cite{9838767} and improves the reliability of the solution. This approach ensures the overall reliability of the solution for STAR-RIS-enabled networks.

Despite the advantages of STAR-RIS, its effectiveness can be limited when deployed in fixed locations, especially in scenarios with mobile users or dynamic environments. By mounting STAR-RIS on UAVs, we can combine the benefits of STAR-RIS technology with the mobility and ability of UAVs to facilitate line-of-sight (LoS) transmission \cite{zhao2022simultaneously}. Employing UAVs for aerial communication can facilitate efficient resource allocation and network optimization \cite{10812765}. This Aerial-STAR approach allows for dynamic positioning of the RIS, potentially improving line-of-sight conditions and adapting to changing user distributions in real-time.


UAV-mounted RIS systems face unique challenges in balancing communication performance with energy constraints. To maximize communication efficiency, we propose a joint optimization framework that considers Aerial-STAR trajectory, active beamforming vectors at the base station, and passive RIS TRCs. To the best of our knowledge this is the first study to include a practical phase shift model of a STAR RIS in UAV communications. The framework also accounts for UAV energy limitations and associated aerodynamic effects of the STAR-RIS.

The key contributions of this work\footnote{This work was supported in part by the Commonwealth Scholarship Commission, U.K.; and in part by the Communications Hub for
Empowering Distributed ClouD Computing Applications and Research (CHEDDAR) funded by U.K. Engineering and Physical Sciences
Research Council (EPSRC) under Grant EP/Y037421/1 and Grant EP/X040518/1.} are as follows:
\begin{itemize}
\item We propose a novel Dual Actor Deep Deterministic Policy Gradient (DA-DDPG) algorithm, which features two separate actor networks designed to effectively manage the complex action space that emerges when implementing practical phase shift designs for STAR-RIS. Specifically, DA-DDPG addresses the challenges posed by the combination of high-dimensional continuous and discrete actions required at the Aerial-STAR.
\item We introduce a new fairness metric called the Harmonic Fairness Index (HFI). Based on harmonic mean, HFI is designed to help achieve a minimum data rate for all users by allowing a more equitable distribution of resources.


\item Simulation results\footnote{All code and research artifacts will be made openly available upon acceptance of the manuscript for publication.} demonstrate that the DA-DDPG algorithm excels by offering specialization for both continuous and discrete actions, surpassing reward accumulation of conventional single-actor DDPG (by 24\%) and DQN (by 97\%) algorithms. The DA-DDPG-based Aerial-STAR offers a 22\% efficiency improvement over conventional RIS and dual Transmit/Reflect (T/R) RIS. Additionally, the optimization of Aerial-STAR 3D trajectory outperforms stationary deployments of STAR RIS and other trajectory configurations. Incorporating HFI in the reward function is shown to provide maximum user fairness among compared approaches.

\end{itemize}
\section{Related Work and Motivation}

\subsubsection{UAV enabled RIS communications} UAVs have seen widespread adoption in wireless networks due to their affordability, exceptional mobility, and ability to facilitate line-of-sight (LoS) transmission \cite{zeng2019accessing}. These properties can be coupled with the capabilities of STAR-RIS to provide better coverage. \cite{su2023joint} proposes installing stationary STAR-RIS on building facades in UAV networks to maximize UAV-User sum rate by optimizing UAV location and STAR RIS TRCs. Similarly, \cite{zhang2022joint} jointly optimizes the beamforming vectors, UAV trajectory, and power allocation to maximize sum rate using fixed STAR-RIS. 

Even though UAVs provide some increased degrees of freedom for communications for stationary STAR-RIS, this does not alleviate the problem of the high probability of NLoS channel between the RIS and mobile users. The concept of UAV-mounted RIS surfaced in \cite{tyrovolas2022energy}, which suggested that a mobile RIS can enhance the communication experience owing to a strong LoS channel. In \cite{do2021aerial}, it was demonstrated that incorporating a UAV-mounted RIS can notably improve outage performance, irrespective of whether the UAV-mounted RIS is stationary or in motion. All of these research endeavors focus on reflect only RIS mounted on UAVs. This concept has recently been extended to UAV mounted STAR-RIS \cite{10458888} which proposes optimization of task offloading, UAV trajectory, RIS TRCs, and transmit power to maximize energy consumption. 

However, in both scenarios, whether the STAR-RIS serves as terrestrial assistance in UAV networks or is mounted directly on UAVs, the common assumption prevails - the independent adjustment of phase-shift coefficients for transmission and reflection. This assumption suggests unrealistic flexibility in arbitrarily varying the corresponding electric and magnetic impedances. This assumption may not apply to passive STAR-RIS configurations, as their feasible electric and magnetic impedance is limited to purely imaginary numbers \cite{liu2022simultaneously} and motivates our work.
\vspace{-3mm}

\subsubsection{Reinforcement Learning (RL) in RIS aided networks} Recent research has shown that RL is capable of intelligently managing RISs and UAV trajectories \cite{liu2020machine,wang2024reinforcement,10376206}. \cite{liu2020machine} introduced a DQN algorithm aimed at enhancing energy efficiency by optimizing both UAV trajectory and passive phase shift RIS coefficients. Similarly, \cite{wang2024reinforcement} proposed a Twin Delayed DDPG (TD3) algorithm to meet the Quality of Service (QoS) constraints by jointly optimizing RIS phase shift and beamforming vectors. However, these approaches suffer from the limitation of providing only 180$^\circ$ coverage. On the other hand, \cite{10376206} and \cite{aung2023deep} employed the STAR RIS for 360$^\circ$ coverage but utilized Double-DQN. The drawback with DQN lies in its inability to output continuous actions suited for RIS phase and UAV trajectories design. Solutions that consider continuous/mixed action space often either set static RIS \cite{9837935} or independent phase shifts \cite{10320337}, both of which suffer the limitations discussed previously. 
\vspace{-3mm}

\subsubsection{RL-based solutions for discrete/continuous action spaces} 
For STAR RIS systems, where optimizing both transmission and reflection coefficients is essential, the action space can be modeled as a combination of continuous and discrete actions (discussed in Sec. \ref{secIV}). Some works have investigated managing continuous and discrete action spaces. For instance, \cite{lowe2017multi} demonstrates that multi-actor centralized critic frameworks can effectively handle coordination challenges and improve learning stability in both continuous and discrete action spaces. \cite{delalleau2019discrete} demonstrated that using specialized output branches for discrete and continuous actions enabled better handling of hybrid action spaces in video game AI than traditional single-output approaches. Likewise, \cite{neunert2020continuous} developed a hybrid actor-critic algorithm with specialized discrete and continuous action outputs for robotic control tasks, achieving superior results over approaches that relied on action space approximations or expert heuristics. To deal with a similar problem, \cite{tavakoli2018action} proposed the Action Branching architectures, showing that distinct network branches for different action types enhance performance in hybrid action spaces. These examples illustrate the potential advantages of using specialized architectural components for STAR-RIS optimization, 
highlighting how they can effectively handle both continuous and discrete actions while maintaining coordination through shared state representations.
\vspace{-4mm}

\subsubsection{Motivation} Analysis of the literature reveals several limitations in existing research efforts. While many efforts focus on employing conventional reflect-only RIS \cite{basharat2021reconfigurable,do2021aerial,zeng2019accessing}, which has a limited service region confined to reflection space, some efforts overcome this limitation by employing STAR-RIS \cite{you2021wireless, niu2021weighted}. However, they often assume pre-installed supporting infrastructure, which may not always be available. Another challenge arises from fixed RIS installations, which may not always have line-of-sight (LoS) channels with mobile users, especially in urban scenarios. UAV-mounted STAR-RIS has been proposed to address these issues \cite{10320337, do2021aerial}, but most studies assume arbitrary phase coefficients, which is impractical for passive RIS. The effective design of coupled transmission and reflection beamforming for STAR-RIS is challenging due to the mutual interdependence. This complexity demands a control scheme that allows integration of continuous and discrete actions — a requirement that existing relevant algorithms, which have been designed for relatively small action spaces \cite{delalleau2019discrete}, cannot easily meet due to the large action dimensions involved in RIS \cite{9837935}. Moreover, treating these actions separately through specialized networks may lead to improved performance. Considering the limitations of existing research and the challenges involved, we propose a DA-DDPG algorithm with two specialized actors, each designed to handle continuous and discrete actions independently. This approach leverages recent advancements in hybrid actor-critic methods to address the high-dimensional action space.

\section{SYSTEM MODEL}\label{secIII}
\subsection{System Description}

We consider a downlink communication network comprising an $M$-antennae base station located at ${\left ({{{x_{b}},{y_{b}},{z_{b}}} }\right)^{T}}$ and Aerial-STAR with $N$ elements located at $\mathbf{T}_s = {\left ({{{x_{s}},{y_{s}},{z_{s}}} }\right)^{T}}$ as shown in Fig.~\ref{system_model}. The STAR-RIS operates in Energy Splitting (ES) mode, dividing the incoming signal energy at each element for transmission and reflection regions. There are ${\mathcal {T}}$ and ${\mathcal {R}}$ single antenna users in the transmit and reflect areas respectively. $J$ is the total number of users such that $J = P + Q $ where $P$ and $Q$ denote the number of $\mathcal{R}$ and $\mathcal{T}$ users, which are located at ${\left ({{{x_{j}},{y_{j}},{z_{j}}} }\right)^{T}}$.

$\beta_{n}^{{\mathcal {R}}}$ and $\beta_{n}^{{\mathcal {T}}}$, $n = 1,2,\ldots,N$ represent the amplitude response for each element for ${\mathcal {R}}$ and ${\mathcal {T}}$ users respectively. Reflect and transmit phase shifts are denoted as $\theta_n^{\mathcal {R}}$, $\theta_n^{\mathcal {T}}$ $\in$ (0,2$\pi$]. The TRCs for each element are denoted as $\beta_{n}^{{\mathcal {T}}} e^{j\theta _{n}^{{\mathcal {R}}}}$ and $\beta_{n}^{R} e^{j\theta _{n}^{R}}$ and need to be optimized for passive beamforming, which is non-trivial as the TRC of each element is controlled by its impedance. For a passive lossless RIS \cite{9838767}, the law of conservation of energy dictates $\beta_{n}^{{\mathcal {R}}}$ = $\sqrt{1-(\beta_{n}^{{\mathcal {T}}})^2}$. The phase shift is coupled by the relationship $\cos({\theta _{n}^{{\mathcal {R}}} - \theta _{n}^{{\mathcal {T}}}}) = 0$ and follows 
\begin{equation}
    | \theta _{n}^{{\mathcal {R}}} - \theta _{n}^{{\mathcal {T}}} | = \pm \frac{\pi}{2}, \label{1}
\end{equation}

The TRC matrices $\boldsymbol{\Phi }$ for an $N$-element STAR-RIS is represented as 
\begin{equation}
\boldsymbol{\Phi}_c = \mathrm{diag}\left( \beta_1^c e^{j\theta_1^c}, \beta_2^c e^{j\theta_2^c}, \ldots, \beta_N^c e^{j\theta_N^c} \right), \quad c \in \{{\mathcal {R}}, {\mathcal {T}}\}
\label{2}
\end{equation}
Thus, the ES-based STAR-RIS can support $N$ multi-paths for both ${\mathcal{R}}$ users and ${\mathcal{T}}$ users. However, as previously mentioned, the advantage of enhanced multi-path gain comes at the cost of limited flexibility in configuring TRCs, which cannot be adjusted arbitrarily or independently.

\subsection{Channel and Signal Model}

In our system, we consider multiple wireless channels that need to be modeled accurately. These include the channel between the Aerial-STAR and base station ($\mathbf{H}_{b}^{s} \in \mathbb{C}^{M \times N}$), the channels from the BS to the ${\mathcal {R}}$ ($\mathbf{H}_b^{\mathcal {R}} \in \mathbb{C}^{{M \times P }}$), BS to ${\mathcal {T}}$ users ($\mathbf{H}_b^{\mathcal {T}} \in \mathbb{C}^{{M \times Q} }$) and channels from the Aerial-STAR to the ${\mathcal {R}}$ ($\mathbf{H}_s^{\mathcal {R}} \in \mathbb{C}^{{N \times P} }$) and ${\mathcal {T}}$ users ($\mathbf{H}_s^{\mathcal {T}} \in \mathbb{C}^{{N \times Q} }$). Here $N$ is the number of elements in the Aerial-STAR, $M$ is the number of BS antennae, $P$ is the number of reflect users, and $Q$ is the number of transmit users.
For each specific user, these channels can be denoted as $\boldsymbol {h}_{b}^u \in {\mathbb {C}^{M \times 1}}$, $\boldsymbol {h}_{s}^u \in {\mathbb {C}^{N \times 1}}$ where $u \in {p,q}$. 

All channels are modeled using a quasi-static block fading approach, in which the fading coefficients stay constant throughout each time slot ($t$) but can change from one slot to the next based on the current position of Aerial-STAR, base station, and users. The channel $\mathbf{H_{b}^{s}}$ between Aerial-STAR and BS is modeled as a Rician fading channel, considering both line-of-sight (LoS) and non-line-of-sight (NLoS) components:
\begin{figure}[t]
    \centering
    \includegraphics[width=\columnwidth]{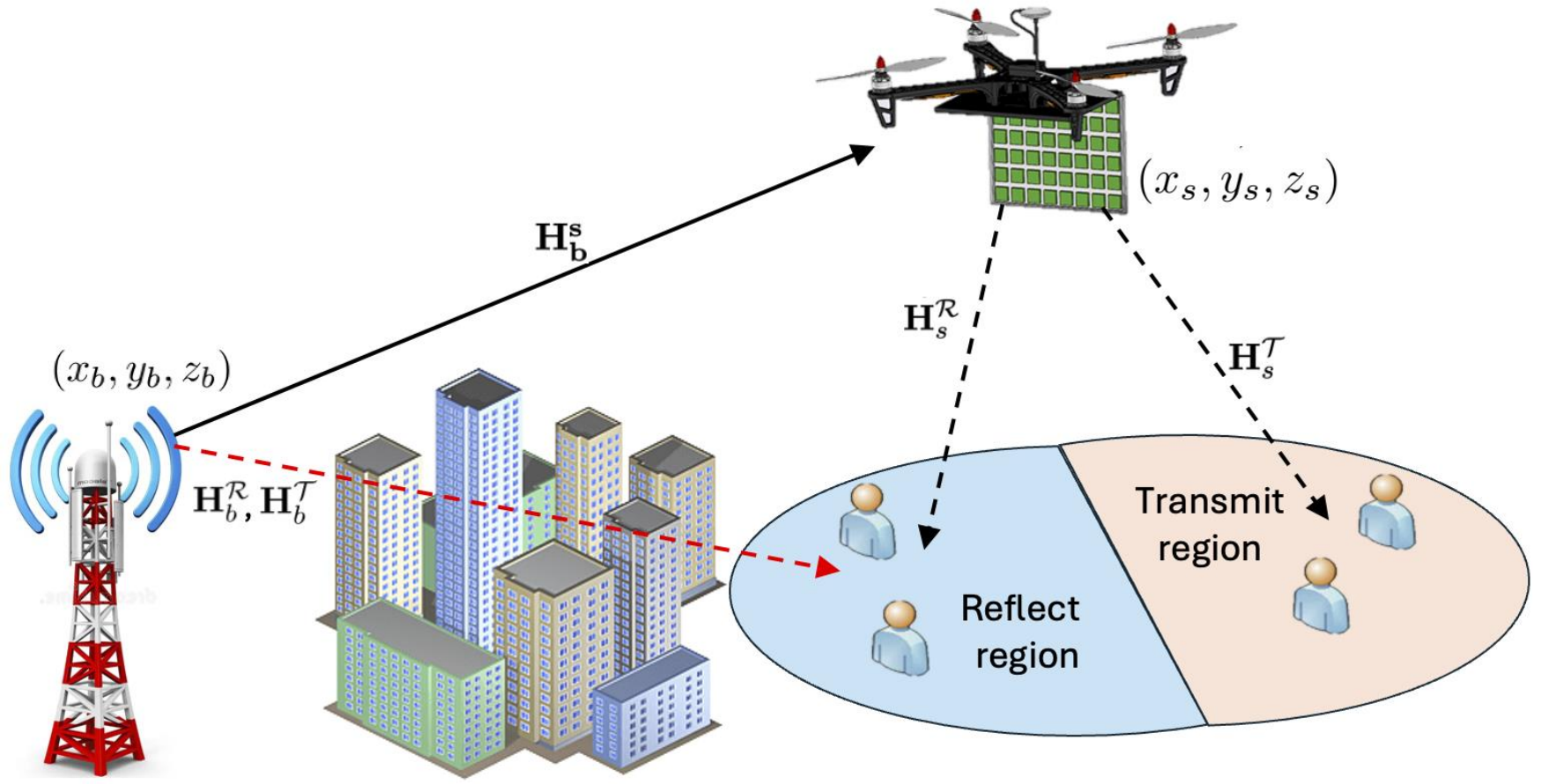}
    \caption{System Model for Aerial-STAR aided network.}
    \label{system_model}
 
\end{figure}
\begin{equation}
    \mathbf{H}_{b,t}^{s} = \sqrt{L(d_{b,t}^{s}, f_c)} \left(\sqrt{\frac{\kappa}{\kappa+1}} \mathbf{G}_{b,\text{LoS}}^s + \sqrt{\frac{1}{\kappa+1}} \mathbf{G}_{b,\text{NLoS}}^s\right)\label{3}
\end{equation}
\noindent where $L(d_{b,t}^{s}, f_c)$ represents the path loss, $\kappa$ is the Rician factor, $\mathbf{G}_\text{b,NLoS}^s \in \mathbb{C}^{N \times M} \sim \mathcal{CN}(0,1)$ represents gain due to scattered paths, and $\mathbf{G}_\text{b,LoS}^s \in \mathbb{C}^{N \times M}$ is LoS link between STAR-RIS and the BS. This is defined as 
\begin{equation}
    \mathbf{G}_{b,\text{LoS}}^s = \mathbf{a}_s(\theta, \phi) \mathbf{a}_b(\psi)^H
    \label{4}
\end{equation}
\noindent where $\mathbf{a}_b \in \mathbb{C}^{M \times 1}$ and $\mathbf{a}_s \in \mathbb{C}^{N \times 1}$ are the antenna array response vectors for the BS and Aerial-STAR, respectively. $\theta$, $\phi$, and $\psi$ represent the azimuth and elevation angles of arrival/departure \cite{lei2023noma}. 
Following \cite{10376206}, for the linear array at the BS and the uniform planar array at the STAR-RIS, the array response vectors are given by:
\begin{equation}
    \mathbf{a}_b[m] = e^{j2\pi(m-1)\Delta_b \sin(\psi)/\lambda}, \quad m = 1, 2, ..., M
    \label{5}
\end{equation}

\begin{equation}
\begin{aligned}
    \mathbf{a}_s[n] = e^{j2\pi(n-1)\Delta_s(\lfloor n/N_x \rfloor \sin(\theta)\sin(\phi)} \\ 
    \times e^{(n-\lfloor n/N_x \rfloor \cdot N_x)\sin(\theta)\cos(\phi))/\lambda}, \\
    \quad n = 1, 2, ..., N
     \label{6}
\end{aligned}
\end{equation}

\noindent where $\Delta_b$ and $\Delta_s$ are the antenna and element spacing, respectively, and $N_x$ is the number of RIS elements in a row.

\subsection{Path Loss Model}
We adopt the 3GPP specification TR 38.901 \cite{3GPP_TR_38.901} urban propagation model to find the path loss $L$. Taking $d$ as the distance between source and destination, and carrier frequency $f_c$, path loss for LoS link is given by 
\begin{equation}
    L_\text{LoS}(f_c,d) = 20.0 \log_{10}(f_c) + 28 + 22\log_{10}(d) 
     \label{7}
\end{equation}

\noindent For NLoS paths:

\begin{equation}
    L_\text{NLoS} = \max[L_\text{NLoS}(f_c,d),L_\text{LoS}(f_c,d)]
     \label{8}
\end{equation}
\begin{equation}
    L_\text{NLoS}(f_c,d) = 36.7 \log_{10}(d) - 0.3(z_s - 1.5) + 26 \log_{10}(f_c) +  22.7
     \label{9}
\end{equation}
All other channels, i.e., $\mathbf{H}_{b,t}^{\mathcal{R}}, \mathbf{H}_{b,t}^{\mathcal{T}}, \mathbf{H}_{s,t}^{\mathcal{R}}, \mathbf{H}_{s,t}^{\mathcal{T}}$,  are modelled as Rayleigh fading ($\kappa=0$) since their LoS propagation is not guaranteed. 

\subsection{Signal Model}
The beamforming matrix at BS is given by $\mathbf{W}_B(t)$. Let $s_j(t)$ and $\mathbf{v}_j(t) \in \mathbb{C}^{M\times1}$ denote the information sequence and the active beamforming vector for user $j$ at the BS. The transmitted signal by BS at time slot $t$ is:
\begin{equation}
    \mathbf{x}_B(t) = \sum_{j=1}^J \mathbf{v}_j(t) s_j(t). \label{10}
\end{equation}
The signal incident at the Aerial-STAR after Gaussian noise, $\mathbf{n}_0$, is given by:
\begin{equation}
    \mathbf{x}_s(t) = \mathbf{H}_{b,t}^s \mathbf{x}_B(t) + \mathbf{n}_0,  \label{11}
\end{equation}
The received signal for reflect user $p$ and transmit user $q$ is:
\begin{equation}
\begin{aligned}
    y_i(t) = [\mathbf{h}_{b,t}^i + \mathbf{h}_{s,t}^i \mathbf{\Phi}_c(t) \mathbf{h}_{b,t}^s] \sum_{i=1}^{P, Q} \mathbf{v}_i(t) s_i(t) + n_0, \\
    (i,c) \in \{{(p,\mathcal {R)}}, {(q,\mathcal {T)}}\}  \label{12}
\end{aligned}
\end{equation}
The signal-to-interference-plus-noise ratio (SINR) with noise power $\sigma^2$ for individual users is given by:

\begin{equation}
\begin{aligned}
    \gamma_i(t) = \frac{|[\mathbf{h}_{b,t}^i + \mathbf{h}_{s,t}^i \mathbf{\Phi}_c(t) \mathbf{h}_{b,t}^s] \mathbf{v}_i(t)|}{|[\mathbf{h}_{b,t}^i + \mathbf{h}_{s,t}^i \mathbf{\Phi}_c(t) \mathbf{h}_{b,t}^s]  \sum_{j=1,j\neq i}^{P,Q} \mathbf{v}_j(t)| + \sigma^2} \\
    \\
    (i,c) \in \{{(p,\mathcal {R)}}, {(q,\mathcal {T)}}\}  \label{13}
\end{aligned}
\end{equation}
The data rate for each user is
\begin{equation}
    R_j(t) = B \log_2(1 + \gamma_j(t))  \label{14}
\end{equation}
\noindent where $B$ is the system bandwidth.

\subsection{Aerial-STAR Energy Model}

Aerial-STAR energy consumption comprises energy required for UAV propulsion and energy required to overcome drag experienced by STAR-RIS. We use the UAV energy model proposed by \cite{8663615} for horizontal motion. The total power consumption for a UAV, \( P_{UAV} \), can be described as:
\begin{equation}
P_{UAV} = P_{blade} + P_{induced} + P_{parasite} + P_{climb}  \label{15}
\end{equation}
\noindent where \( P_{blade} \) refers to the blade profile power, which compensates for viscous drag losses in the rotor blades and \( P_{induced} \) is the induced power, accounting for the energy required to overcome the induced drag on the rotor blades. Additionally, \( P_{parasite} \) is the parasitic power needed to overcome drag on the UAV's fuselage, while \( P_{climb} \) is the power needed to change the potential energy during ascent or descent. These terms are mathematically represented as:

\begin{subequations} \label{16}
\begin{align}
& P_{blade} = \underbrace {\frac {\rho \delta A s (r\Omega)^3}{8}}_{\text{hovering condition}} \times \underbrace {\left(1 + \frac {3V^2}{(r\Omega)^2}\right)}_{\text{velocity correction factor}}, \tag{16a}\label{16a} \\
& P_{induced} = \underbrace {\frac {T^{3/2}}{2 \rho A}}_{\text{hovering condition}} \times \underbrace {\left(\sqrt{\frac{V^4}{4v_{i}^{4}} + 1} - \frac{V^2}{2v_{i}^{2}}\right)}_{\text{velocity correction factor}}, \tag{16b}\label{16b} \\
& P_{parasite} = \frac {\rho s A d_0 V^3}{2}, \quad P_{climb} = \pm T \dot{h}, \tag{16c}\label{16c}
\end{align}
\end{subequations}

In these expressions, \( \rho \), \( \delta \), \( \Omega \), \( A \), \( s \), and \( r \) represent the air density, profile drag coefficient, angular velocity of the rotor, rotor area, solidity, and rotor radius, respectively. The product \( r\Omega \) represents the blade tip speed. Meanwhile, \( V \), \( T \), \( v_i \), and \( d_0 \) stand for the airspeed, thrust, induced velocity, and fuselage drag coefficient, respectively. $\dot{h}$ is the rate of climb or descend. It should be noted that for constant climb / descent rates, $T$ equates to the weight of the UAV.

Our model addresses a critical gap in the existing literature by specifically incorporating the energy required to overcome drag caused by the STAR-RIS. This factor has frequently been neglected in earlier studies when varying the number of RIS elements to compute energy efficiency~\cite{yang2024energy,tariq2023reinforcement,10594363, 10540672}.

Given that the RIS operates at $f_c$, the wavelength is $\lambda = c/f_c$. Typical RIS element spacing can be $\lambda/\mu$ where  $\mu \in (2, 8)$ \cite{tang2020wireless}. Hence, for a square RIS with $N_x$ elements in each row, the surface area is given by
\begin{equation}
A_{RIS}= (N_x-1)^2 \times\lambda^2/\mu^2   \label{17}
\end{equation}
\noindent The power required to overcome RIS drag would be
\begin{equation}
P_{RIS-drag} = \frac{1}{2} \rho A_{RIS} C_d V_\perp^3  \label{18}
\end{equation}

where $\rho$ is the air density, $C_d$ is the RIS drag coefficient (typically 2.1 for near square surfaces \cite{gerhart2016munson}), and $V_\perp$ is the component of velocity perpendicular to RIS plane. 

Then total power for time slot $t$ is given by 

\begin{equation}
P_{total}(t) = P_{UAV} + P_{RIS-drag}  \label{19}
\end{equation}

\begin{figure}[t]
    \centering
    \includegraphics[width=\columnwidth]{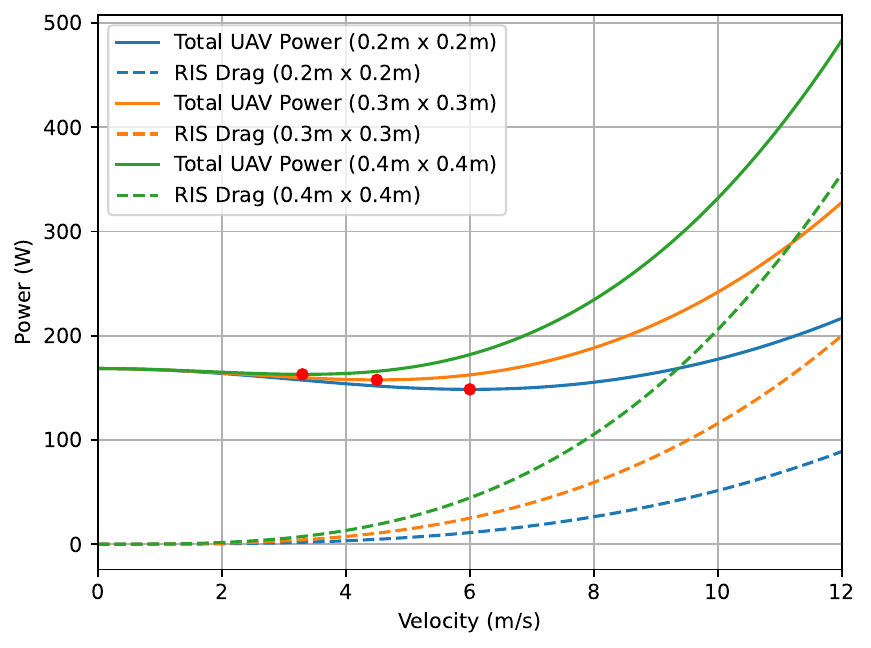}
    \caption{Larger RIS surfaces result in higher drag, leading to increased power consumption, especially at higher velocities, with red markers indicating the most energy-efficient velocities for each case. The figure shows that it is important to include aerodynamic drag due to Aerial-STAR, especially at high values of area and speed.}
    \label{RIS_size_vs_energy}
    {\vspace{-8mm}}
\end{figure}

The communication efficiency of system at time slot $t$ is defined as 
\begin{equation}
\eta_{EE}(t) = \frac{\sum_{j=1}^{J}R_j(t)}{P_{total}(t)}  \label{20}
\end{equation}
By incorporating the energy consumption due to RIS-induced drag and using a detailed inter-element spacing model, we present a more accurate representation of UAV energy dynamics in RIS-enabled systems\footnote{Wind can have a non-trivial effect on the RIS surface and may change the UAV attitude. To recover to the original position, the UAV needs extra power which is beyond the scope of this paper. Similarly, the effect of different RIS shapes on total drag energy should also be studied in future works.}. It can be noted from these equations that $A_{RIS}$, and consequentially $P_{RIS-drag}$, grows quadratically with $N_x$. Therefore, a higher number of RIS elements might result in a higher sum rate of the system but would not necessarily improve the energy efficiency of the system due to higher energy consumption as shown in Fig.~\ref{RIS_size_vs_energy}. This model may be valuable for optimizing UAV-RIS configurations and play a role in addressing practical deployment challenges.

\subsection{Problem Formulation}
We aim to maximize the energy efficiency of the system by jointly optimizing UAV trajectory, $\mathbf{T}$, STAR-RIS TRCs, $\mathbf{\Phi}$, and beamforming vectors at BS, $ \mathbf{W}_B$. This problem is formulated as \vspace{-4mm}

\begin{subequations} 
\begin{align}
&\max _{ \mathbf{T}, \mathbf{\Phi}, \mathbf{W}_B} \sum _{t=1}^{T}   \mathbf{\eta}_{EE}(t), \label{21a} \\
&\quad ~\textrm{s.t.}\,\,-\pi \leq \theta _{n,t}^\mathcal{T} \le \pi, \quad \forall n,~\forall t, \label{21b}\\
&\hphantom {\quad ~\textrm{s.t.}~} -\pi \leq \theta _{n,t}^\mathcal{R} \le \pi,\quad \forall n,~\forall t, \label{21c}\\
&\hphantom {\quad ~\textrm{s.t.}~} R_{k,t} \geq R_{\text {QoS}},\quad \forall k, ~\forall t,  \label{21d}\\
&\hphantom {\quad ~\textrm{s.t.}~}  \beta _{n,t}^\mathcal{R},\beta _{n,t}^\mathcal{T} \in [0,1],\quad \forall n,~\forall t, \label{21e}\\
&\hphantom {\quad ~\textrm{s.t.}~} \cos (\theta _{\mathcal {R},n,t} - \theta _{\mathcal {T},n,t}) = 0,  \label{21f}\\
&\hphantom {\quad ~\textrm{s.t.}~} \parallel\mathbf {w}^{2}_{k,t}\parallel\leq P_{\text {max}},  \label{21g}\\
&\hphantom {\quad ~\textrm{s.t.}~} \parallel \mathbf{T}_{s,t+1} - \mathbf{T}_{s,t}\parallel \leq V_{\text {max},}\label{21h}
\end{align}
\end{subequations} 

\noindent where \eqref{21a} represents the objective function to maximize energy efficiency, \eqref{21b} and \eqref{21c} is the range constraint for TRC phase shifts, and \eqref{21d} represents the minimum data rate constraint. \eqref{21e} limits the amplitude response to a legitimate range, \eqref{21f} lays the coupling constraint between transmit and reflect co-efficients, \eqref{21g} is the maximum power constraint, and \eqref{21h} constrains maximum speed of the Aerial-STAR system. 

It should be noted that the coupling constraint \eqref{21f} adds an extra layer of challenge in jointly optimizing TRCs. The requirement implies that only one, either $\theta_n^\mathcal{R}$ or $\theta_n^\mathcal{T}$, can have continuous domain. The other would deviate from the first by $+\pi/2$ or $-\pi/2$. Therefore, the binary selection for phase shift could be handled more efficiently with discrete actions instead of continuous actions.

\vspace{-1mm}
\section{Proposed Solution}\label{secIV}
To address the optimization challenge in our STAR-RIS system, we formulate the problem within the framework of a DRL. Specifically, we model the transmission period as a Markov decision process (MDP), enabling the application of deep reinforcement learning techniques. In this formulation, each time slot t $\in$ T represents a discrete step in the process. 

\vspace{-3mm}
\subsubsection{State and Action} The agent in our case is a network controller that could be either placed onboard the Aerial-STAR to reduce latency while taking actions in real-time, or the base station which has a LoS link with the Aerial-STAR and can help reduce the computation at the UAV. The agent observes the current state $\mathbf{s}_t \in \mathbf{S}$, where $\mathbf{S}$ denotes the state space. Based on this state observation, the agent selects an action $\mathbf{a}_t \in \mathbf{A}$ from the action space $\mathbf{A}$. for each $t$,  $\mathbf{s}_t$ has dimension $(3+ 2MN + 2MJ + 2NJ)$ and contains information about Aerial-STAR position and channel information as given by 
\begin{equation}
    \mathbf{s}_t = \{\mathbf{T}_{s,t}, \mathbf{H}_{b,t}^{s},  \mathbf{H}_{b,t}^{\mathcal{R}}, \mathbf{H}_{b,t}^{\mathcal{T}}, \mathbf{H}_{s,t}^{\mathcal{R}}, \mathbf{H}_{s,t}^{\mathcal{T}}\}\label{22}
\end{equation}

The DA-DDPG algorithm utilizes two separate actor networks: one for continuous actions and one for discrete actions. These networks operate in parallel but are trained jointly through the critic network. The state is input to continuous and discrete actor networks which output continuous $(\mathbf{a}_t^c)$ and discrete $(\mathbf{a}_t^d)$ actions
\begin{subequations} 
\begin{align}
& \mathbf{a}_t^c = \{\mathbf{a}_t^\mathbf{T}, \mathbf{a}_t^{\mathbf{\Phi}_\mathcal{R}}, \mathbf{a}_t^{\boldsymbol{\beta}}, \mathbf{a}_t^{\mathbf{w}_t }\} \label{23a} \\
& \mathbf{a}_t^d = \{\mathbf{a}_t^{\mathbf{\Phi}_\mathcal{T}}\} \label{23b}
\end{align}
\end{subequations}

\noindent respectively. $\mathbf{a}_t^c$ is of size $3+2N+2MJ$, while $\mathbf{a}_t^d$ is of size $N$. 

Upon executing the chosen action $\mathbf{a}_t$, the system transitions to a new state $\mathbf{s}_{t+1}$ according to the underlying channel dynamics and user mobility patterns. This transition can be represented as the tuple $(\mathbf{s}_t, \mathbf{a}_t, \textbf{r}_t, \mathbf{s}_{t+1})$, where $\textbf{r}_t$ denotes the reward received as a result of the action.

\begin{figure*}[!t]
\vspace*{-4mm}
\centering
\includegraphics[width=1.8\columnwidth]{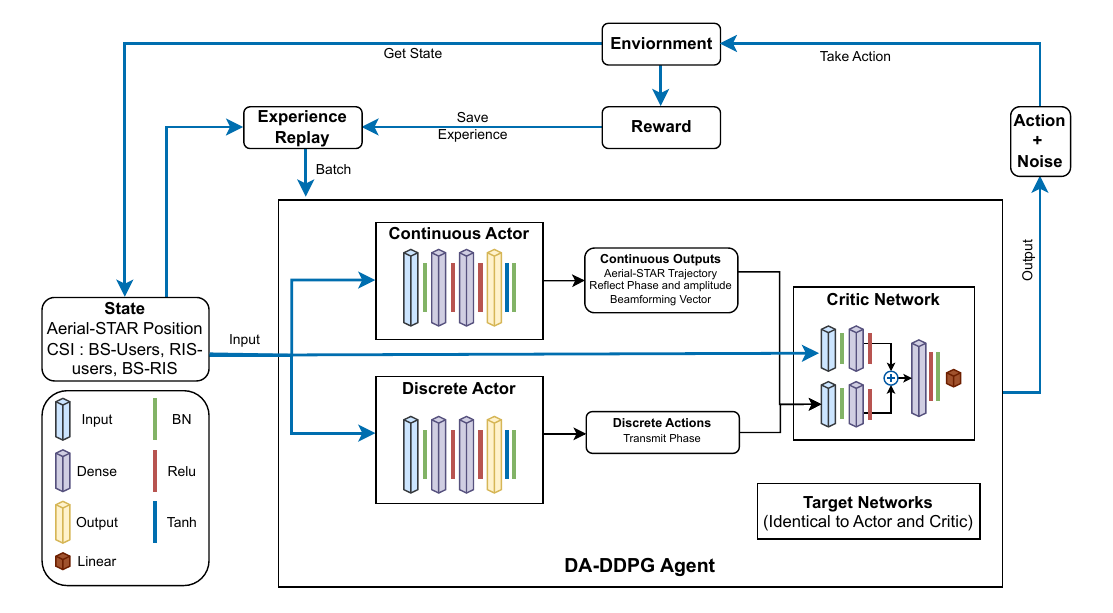}%
    \caption{Architecture of proposed DA-DDPG algorithm with separate actor networks for continuous and discrete outputs}
    \label{DADDPG_architecture}
\vspace{-3mm}
\end{figure*}

\subsection{DA-DDPG Learning Process}

The DA-DDPG algorithm aims to maximize the expected cumulative reward from each state $\mathbf{s}_t$. This is achieved by determining the optimal action ${\{\mathbf{a}_t^c,\mathbf{a}_t^d\}}$ for each state $\mathbf{s}_t$. Specifically, the algorithm seeks to find the actions that maximize the expected return, which is the sum of the discounted future rewards starting from state $\mathbf{s}_t$. Mathematically, the goal is to maximize the value function $Q^\mu(\mathbf{s}_t, {\{\mathbf{a}_t^c,\mathbf{a}_t^d\}})$, which represents the expected cumulative reward for taking action ${\{\mathbf{a}_t^c,\mathbf{a}_t^d\}}$ in state $\mathbf{s}_t$ and following the optimal policy thereafter using the bellman equation:
\begin{equation}
\begin{aligned}
Q^\mu(\mathbf{s}_t, \{\mathbf{a}_t^c, \mathbf{a}_t^d\}) 
&= \mathbb{E}\left[ r(\mathbf{s}_t, \{\mathbf{a}_t^c, \mathbf{a}_t^d\}) \right. \\
& \quad + \zeta Q^\mu(\mathbf{s}_{t+1}, \{\mathbf{a}_{t+1}^c, \mathbf{a}_{t+1}^d\}) ]\label{24}
\end{aligned}
\end{equation}

\noindent where $\zeta$ is the discount factor in the range [0,1], $\mu$ represents the action policy that maps $\mathbf{s}_t$ to $\mathbf{a}_t$, and $r(\cdot)$ is the reward received. The optimal action, $\mathbf{a}_t^*$ is represented as 

\begin{equation}
    \{\mathbf{a}_{t}^{*c},\mathbf{a}_{t}^{*d|}\} = \arg\max_{a_t \in \mathbf{A}}Q^\mu(\mathbf{s}_t, \{\mathbf{a}_t^c,\mathbf{a}_t^d\}) \label{25}
\end{equation}

Our approach uses two parameterized actor functions $\mu^c(\mathbf{s|\omega}_{\mu}^c)$ and $\mu^d(\mathbf{s|\omega}_{\mu}^d)$, along with a critic function approximator $Q(\mathbf{s}, \{\mathbf{a}^c,\mathbf{a}^d\}|\omega_Q)$. We train the DA-DDPG agent by minimizing the following loss function:

\begin{equation}
L(\mathbf{\omega}_Q) = \frac{1}{e}\sum_e [y_t - Q(\mathbf{s}_t, {\{\mathbf{a}_t^c,\mathbf{a}_t^d\}}|\omega_{Q,t})]^2,\label{26}
\end{equation}

\noindent where $L(\omega_Q)$ is the loss function for the critic network, $\omega_Q$ are the parameters of the critic network, and $e$ represents the number of sampled transitions.

The target Q-value $y_t$ is provided by the target network:

\begin{equation}
\begin{aligned}
y_t 
&= r(\mathbf{s}_t, {\{\mathbf{a}_t^c,\mathbf{a}_t^d\}}) \\
& \quad + \zeta Q[\mathbf{s}_t, \mu'^c(\mathbf{s}_t|\omega_{\mu',t}^c), \mu'^d(\mathbf{s}_t|\omega_{\mu',t}^d)|\omega_{Q'}^t ]\label{27}
\end{aligned}
\end{equation}

\noindent where $\omega_{\mu'}^c, \omega_{\mu'}^d, \omega_{Q'}$ are the parameters of the continuous and discrete target actors and critic networks, respectively.

We train the actor networks using policy gradients derived from the critic network per \eqref{28}, where \(\nabla_{\omega_{\mu}^c} J\) and \(\nabla_{\omega_{\mu}^d} J\) are the policy gradients for the continuous actor \(\mu^c\) and the discrete actor \(\mu^d\), respectively, and \(\mathbf{s}_e\) and \(\mathbf{a}_e\) represent the sampled states and actions from the experience replay buffer.

\begin{figure*}[t]
\begin{equation}
\begin{aligned}
\nabla_{\omega_{\mu}^c} J &= \frac{1}{e}\sum_e \nabla_{\mathbf{a}} Q(\mathbf{s}_e, \{\mathbf{a}_e^c, \mathbf{a}_e^d\}|\omega_Q)|_{\mathbf{s}_e=\mathbf{s}_t, \mathbf{a}_e=\{\mu^c(\mathbf{s}_t|\omega_{\mu}^c), \mathbf{a}_e^d\}} \nabla_{\omega_{\mu}^c} \mu^c(\mathbf{s}_e|\omega_{\mu}^c)|_{\mathbf{s}_e=\mathbf{s}_t} \\
\nabla_{\omega_{\mu}^d} J &= \frac{1}{e}\sum_e \nabla_{\mathbf{a}} Q(\mathbf{s}_e, \{\mathbf{a}_e^c, \mathbf{a}_e^d\}|\omega_Q)|_{\mathbf{s}_e=\mathbf{s}_t, \mathbf{a}_e=\{\mathbf{a}_e^c, \mu^d(\mathbf{s}_t|\omega_{\mu}^d)\}} \nabla_{\omega_{\mu}^d} \mu^d(\mathbf{s}_e|\omega_{\mu}^d)|_{\mathbf{s}_e=\mathbf{s}_t} \label{28}
\end{aligned}
\end{equation}
\vspace{-6mm}
\end{figure*}

\subsection{Action Space and Policy Mapping}
The action at time $t$ for our continuous actor is formulated as:
\begin{equation}
\mathbf{a}^{c}_t = \mu^c(s_t|\omega_{\mu,t}^c) + \mathcal{N}_{\text{OU}}(0, \sigma^2), \label{29}
\end{equation}

\noindent where $\mathcal{N}_{\text{OU}}(0, \sigma^2)$ is the OU noise \cite{abraham2021ergodic} with mean $0$ and variance $\sigma^2$, and added to encourage exploration in the continuous action space. The noise is decayed to favor exploitation as the training progresses.  $\mathbf{a}_t^c$ is then mapped to actions \eqref{23a} to be taken by the agent. Aerial-STAR trajectory includes  $\{ V,\theta_\mathbf{T},\varphi_\mathbf{T}\}$, which is the travel distance with constraint \eqref{21g}, angle of motion in elevation and azimuth such that $-\pi/2 \le {\theta_\mathbf{T}} \le \pi/2$ and $0 \le {\varphi_\mathbf{T}} \le 2\pi$. The mapping of other continuous outputs in $\mathbf{a}_t^{\mathbf{\Phi}_\mathcal{R}}, \mathbf{a}_t^{\boldsymbol{\beta}}, \mathbf{a}_t^{\mathbf{w}_t }$ is done as per \eqref{21c}, \eqref{21e} and \eqref{21g}, respectively. 

Similarly, the discrete actor policy outputs the actions, where the thresholding process maps the continuous output to discrete values \cite{9837935}:

\begin{equation}
\mathbf{a}_{t,discrete} = \begin{cases}
\frac{\pi}{2} & \text{if } \mu^d(s_t|\omega_{\mu,t}^d)  > 0 \\
-\frac{\pi}{2} & \text{if } \mu^d(s_t|\omega_{\mu,t}^d)  \leq 0
\end{cases}\label{30}
\end{equation}

Then we utilize an epsilon-greedy strategy for exploration.
\begin{equation}
\mathbf{a}_t^d = 
\begin{cases} 
\text{Random from } \{\frac{\pi}{2}, -\frac{\pi}{2}\} & \text{with probability } \epsilon \\
\text{Thresholded } \mathbf{a}_{t,discrete} & \text{with probability } 1 - \epsilon 
\end{cases}\label{31}
\end{equation}

The exploration rate $\epsilon$ is decayed over time to reduce exploration as the agent learns. Then for STAR-RIS element $n$ with discrete action $a_{n,t}^d \in \mathbf{a}_t^d$ at time $t$, 
\begin{equation}
    \theta _{n,t}^\mathcal{T} = \theta _{n,t}^\mathcal{R} + a_{n,t}^d.
    \label{32}
\end{equation}

\subsection{Reward Structure}

\subsubsection{Enhancing Fairness through Harmonic Fairness Index based reward function}

In wireless networks, fairness metrics are commonly used to measure how evenly resources are distributed. The most commonly used metric is, Jain's Fairness Index (JFI), which was introduced in \cite{Jain_Fairness_Index} and is given by
\begin{equation}
    JFI = \frac{(\sum_{j=1}^J R_j)^2}{J \sum_{j=1}^n (R_j)^2}.
    \label{eq:jfi}
\end{equation}

However, JFI's sensitivity to fairness is limited when users have a high degree of throughput disparity. Due to its reliance on the arithmetic mean, JFI can yields relatively high fairness scores even in scenarios where significant performance gaps exist between users, which could be disadvantageous for users with lower throughput \cite{hossfeld2018new}. In RIS-assisted networks, where user throughput can vary greatly due to obstacles, UAV distance, and interference, JFI’s leniency toward disparity can result in suboptimal resource allocation for users in challenging conditions.

To address these limitations, we introduce the Harmonic Fairness Index (HFI) as a novel metric for evaluating fairness in resource allocation, particularly in UAV-RIS assisted networks. The harmonic mean, one of the Pythagorean mean \cite{chakrabarty2021four}, is calculated as the inverse average of the reciprocal sum. We define HFI as the ratio of the harmonic mean to the arithmetic mean of user throughputs. 
\begin{equation}
    HFI = \frac{\text{Harmonic Mean}}{\text{Arithmetic Mean}} = \frac{J}{\sum_{j=1}^J \frac{1}{R_j}} \cdot \frac{1}{\frac{1}{J}\sum_{i=1}^J R_j}
    \label{eq:hfi}
\end{equation}

\subsubsection{HFI comparison with JFI}

While HFI and JFI provide a measure of fairness between 0 and 1, with 1 indicating perfect fairness, HFI offers enhanced sensitivity to under-served users. This is due to the harmonic mean's property of being strongly influenced by small values in a dataset. JFI tends to be more lenient in scenarios with high throughput disparity.

\begin{figure}[b]
    \centering
    \includegraphics[width=\columnwidth]{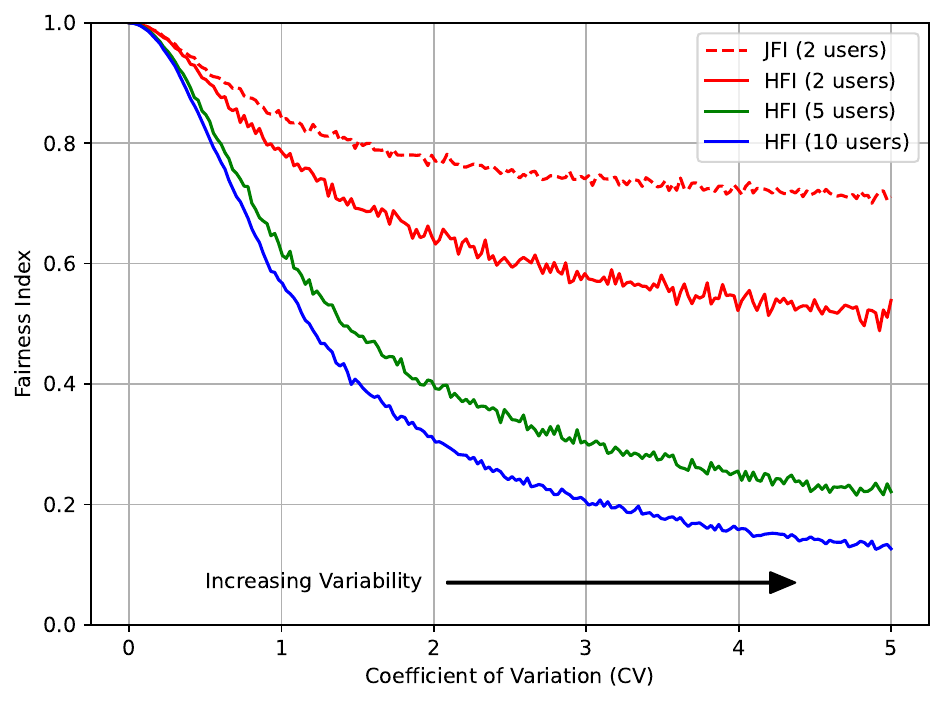}
    \caption{Comparison of HFI and JFI for varying Coefficients of Variation. HFI outputs a lower fairness index for the same users as compared to JFI. Moreover, for the same CV, the penalty of HFI increases as the number of users increases.}
    \label{hfi_vs_jfi}
\end{figure}

Fig. \ref{hfi_vs_jfi} illustrates the behavior of HFI and JFI across different Coefficients of Variation (CV). The CV is the ratio of the throughput standard deviation to its mean, which gives us a normalized measure of throughput disparity. As CV increases, both indices decrease, with HFI exhibiting a steeper decline. This demonstrates HFI's higher sensitivity to throughput variations, making it more suitable for scenarios prioritizing a minimum data rate for all users. HFI's sensitivity increases with user count, promoting fairer resource allocation in larger networks.\footnote{It's worth noting that while our focus is on communication systems, HFI may have a broader applicability in various fields where measuring fairness or equality is crucial, such as economic wealth distribution, resource allocation in computing systems, or even in social policy evaluations.}

In UAV-RIS systems, where network conditions can vary significantly due to factors like UAV distance, obstacles, and interference, HFI's enhanced sensitivity is particularly beneficial. Incorporating HFI into optimization objectives helps in fair resource allocation while maximizing overall throughput. HFI's rapid decrease with increasing CV encourages more aggressive load balancing and resource reallocation in high-disparity scenarios, leading to more dynamic UAV positioning and RIS configuration strategies. This allows HFI to provide a more equitable user experience across the network as demonstrated in Sec.\ref{secV}.

The reward function for our UAV-RIS system is designed to maximize user fairness and sum rate and minimize total power consumption.
\begin{equation}
    \mathbf{r}_t = \alpha \cdot \text{HFI} \cdot \sum_{j=1}^J R_{j,t} - \beta \cdot P_{total,t}
    \label{eq:reward_function}
\end{equation}
where $\alpha$ and $\beta$ are scaling factors. The first term promotes both fairness amongst users and high overall sum rate. By multiplying system sum rate with HFI, which is sensitive to low-performing users, we ensure that the system not only maximizes total data transfer in each time interval but also maintains equitable service (i.e. maintains a comparable data rate) across all users. The second term introduces a penalty for excessive power consumption, nudging the system towards energy-efficient operations. By adjusting the ratio $\beta / \alpha$, we can fine-tune the trade-off between fairness, sum rate, and energy consumption, allowing for flexible adaptation to various operational priorities and constraints in Aerial-STAR systems. For our experiments, we chose values for $\beta / \alpha$ between 0.1-0.2 to prioritize better network performance over saving energy. The optimal ratio, however, may vary depending on the specific scenarios (e.g., number and distribution of users, UAV energy constraints) and priorities (data rate vs energy) of the system.

\subsection{DA-DDPG Architecture}

The structure of the DA-DDPG network, consisting of two actor networks (continuous and discrete) and one critic network, has been carefully designed in line with the coupling relationship between discrete and continuous actions discussed in Problem Formulation (Sec. \ref{secIII}). The actor networks include an input layer accepting the state input with batch normalization, followed by dense layers, each using ReLU activation functions. The output layer employs a tanh activation function to produce continuous actions. The discrete actor network shares a similar structure with the continuous actor, but its output layer corresponds to the number of discrete actions. The critic network is structured with two input branches: one for state and one for actions, each with a batch normalization layer as shown in Fig.\ref{DADDPG_architecture}. The state branch consists of dense layers and ReLU activations. The action input, combining both continuous and discrete actions, passes through a dense layer with ReLU activation. These branches are then concatenated and followed by dense layers with ReLU activations. The size of the hidden layers in all networks is determined based on the state and action dimensions, which may vary depending on the specific problem domain. 
\begin{algorithm}[t]
\caption{DA-DDPG Algorithm}\label{alg:daddpg}
\SetInd{0.2em}{0.8em}
Initialize environment\\
Initialize DA-DDPG with the actor networks $\omega_{\mu}^c$, $\omega_{\mu}^d$, critic network $\omega_Q$, target actor networks $\omega_{\mu'}^c$, $\omega_{\mu'}^d$, target critic network $\omega_{Q'}$\\
Initialize experience replay\\
\For{episode=1 $\rightarrow$ maximum episodes}
    {Set environment to $\mathbf{s}_{t=0}$\\
    \For{$t=0 \rightarrow T$}
        {Observe $\mathbf{s}_t$\\
        Choose actions $\{\mathbf{a}_t^c, \mathbf{a}_t^d\}$ as per \eqref{29},\eqref{23b}\\
        Execute $\{\mathbf{a}_t^c, \mathbf{a}_t^d\}$ \\
        Observe $\mathbf{r}_t$ and next state $\mathbf{s}_{t+1}$ \\
        Store transition $\{\mathbf{s}_t, \{\mathbf{a}_t^c, \mathbf{a}_t^d\}, \mathbf{r}_t, \mathbf{s}_{t+1}\}$ in experience replay \\
        Randomly sample transitions $\{\mathbf{s}_e, \{\mathbf{a}_e^c, \mathbf{a}_e^d\}, \mathbf{r}_e, \mathbf{s}_{e+1}\}$ from experience replay \\
        Set target as per \eqref{27}\\
        Update critic $Q(\mathbf{s}_t, \omega_{Q,t})$ with the minimizing loss \eqref{26}\\
        
        Update actor policies $\mu_c(\mathbf{s}_t, \omega_{\mu,c}^t)$ and $\mu_d(\mathbf{s}_t, \omega_{\mu,d}^t)$ using \eqref{28}\\
        Update the target networks: $\omega_{\mu',t}^i \leftarrow (1-\tau)\omega_{\mu',t}^i + \tau\omega_{\mu',t}^i$, for $i \in \{c,d\}$ and $\omega_{Q'}^t \leftarrow (1-\tau)\omega_{Q'}^t + \tau\omega_{Q'}^t$ \\
        $\mathbf{s}_t \leftarrow \mathbf{s}_{t+1}$
        
        }
    }
\end{algorithm}

\subsection{Complexity Analysis}

The complexity analysis involves calculating the floating point operations required for a pass through these networks. We assume $I$ layers each with $\theta_i$ nodes in the actors and critic networks. $\theta_t, \theta_r$ and $\theta_b$ denote number of nodes in tanh, Relu and BN layers, which require 6, 1 and 5 floating point operations, respectively \cite{8731635}. The complexity of our discrete actor is given as 

\begin{equation}
    \mathcal{C}^d = \mathcal{O}\left(\sum_{i=0}^{I_d} \theta_{i}^d \cdot \theta_{i+1}^d+ 6\theta_t^d +\theta_r^d +5\theta_b^d\right)
\end{equation}
Similarly, $\mathcal{C}^c$ and $\mathcal{C}^Q$ denote the complexity of the continuous actor and critic network, respectively. Both actors take the same input and operate in parallel (independent forward passes and gradient calculation) and so the complexity is dominated by the more complex network. Hence the complexity of dual actors is given by $\mathcal{O}(\max(\mathcal{C}^c,\mathcal{C}^d))$. Hence, the total complexity is given by $\mathcal{O}(\max(\mathcal{C}^c,\mathcal{C}^d) + \mathcal{C}^Q)$. It should be noted that $\mathcal{C}^c$ would almost always be higher than $\mathcal{C}^d$ due to the large action space of the continuous actor network.

For a transmission period with $e$ epochs and $T$ timesteps, the overall complexity can be calculated as 
\begin{equation}
    \mathcal{O}(T \cdot e \cdot (\max(\mathcal{C}^c,\mathcal{C}^d) + \mathcal{C}^Q)).
\end{equation}

It should be noted that in a conventional DDPG, with a single network, the action space would comprise all  $\mathbf{a}_t^c + \mathbf{a}_t^d$ actions. With the same actor architecture as that of DA-DDPG, the complexity of conventional DDPG actor would be higher since $\max(|\mathbf{a}^c|,|\mathbf{a}^d|) \leq$ ($|\mathbf{a}^c| + |\mathbf{a}^d|$).

\section{Results and Analysis}\label{secV}

We perform simulations on an Apple M1 Pro with 3.2 GHz CPU, 16 GB RAM, macOS Ventura 13.2.1. TensorFlow 2.5 is used in JupyterLab on an M1 Pro GPU with 16 cores. We assume that users are randomly distributed on the  $\mathcal{R}$ or $\mathcal{T}$ regions. The Base station is 1 km away from the Aerial-STAR, which operates by staying between the $\mathcal{R}$ and $\mathcal{T}$ regions. We assume that each episode lasts for 30 seconds and the wireless channels update once each second. Adam optimizer is used for training and a default learning rate is set as 0.0001. The actor networks have 2 ReLU and 1 Tanh activations with 256-512 neurons each (depending upon RIS elements $N$), whereas the critic network has 3 dense layers with ReLU activations along with concatenation layer. Table \ref{Simulation Parameters} lists values of all important parameters. Most notations and values used for UAV parameters are referred from \cite{8663615,bramwell2001bramwell}.

\begin{table*}[t]
\caption{Simulation Parameters}
\centering
\begin{tabular}{|c| c| c| c| c| c| } 
    \hline
    \textbf{Parameter} & \textbf{Notation} & \textbf{Quantity} & \textbf{Parameter} & \textbf{Notation} & \textbf{Quantity}\\
    \hline
    Carrier frequency& $f_c$ & 5 GHz & Batch size & $e$ & 64 \\  
    \hline
    STAR RIS elements & $N$ & 16 & Total Rotor Disc Area & $A_r$ & 0.503 m$^2$\\  
    \hline
    BS Antennae & $M$ & 4 & Blade Angular Velocity & $\Omega$ & 300 rad/sec\\  
    \hline
    Channel bandwidth & B & 1 MHz & Profile drag co-efficient & $\delta $ & 0.012\\  
    \hline
    Users& $J$ & 4 & Fuselage drag ratio & $d_0$ & 0.6\\  
    \hline
    Noise density & $\sigma$ & -95 dBm/Hz &  Induced velocity of rotor & $v_i$ & 4.028 m/s\\  
    \hline
    Transmission power& $P_{max}$ & 29 dBm & Rotor Solidity & $s$ & 0.05\\  
    \hline
    Rician factor & $\kappa$ & 5 & Blade tip speed &  $r\cdot\Omega$  & 120 m/s\\  
    \hline
    QoS requirement & $R_{QoS}$ & 200kb/s& Aerial-STAR weight& $W$ & 25 N \\  
    \hline
    Learning Rate & lr & 0.0005 & Air Density & $\rho$ & 1.225 kgm$^{-3}$\\
    \hline
    Discount Factor & $\zeta$ & 1 & Max Aerial-STAR speed & $V_{max}$ & 10 m/s \\  
    \hline
    \end{tabular}
    \label{Simulation Parameters}
\end{table*}

\vspace{-5mm}
\begin{figure}[b]
    \centering
  \includegraphics[scale=0.55]{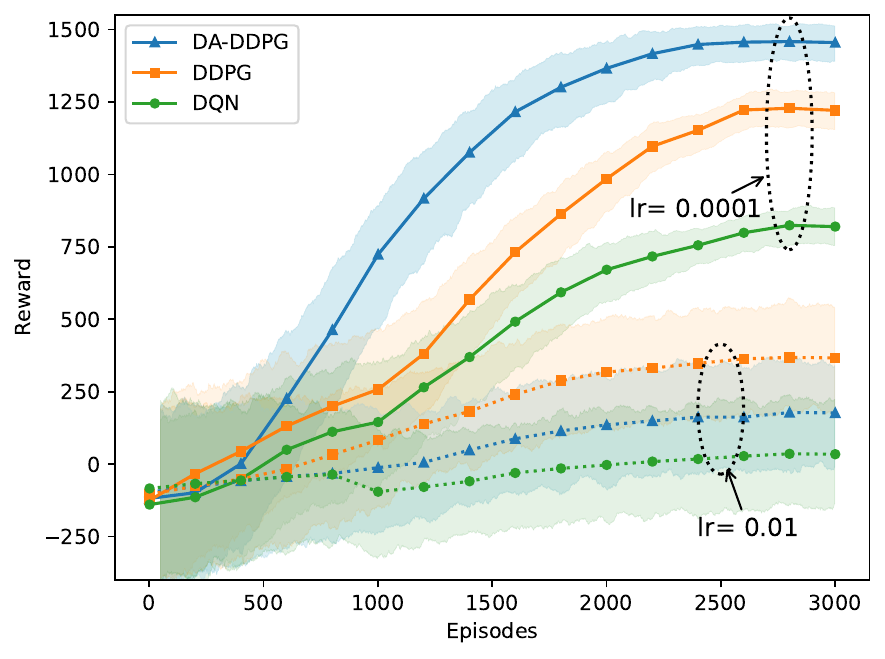}
  \caption{Comparative reward plots for DA-DDPG, DDPG and DQN at different settings of learning rates.}
  \label{reward_vs_architecture}

\end{figure}

\subsubsection{DA-DDPG comparison with relevant RL algorithms} Fig. \ref{reward_vs_architecture} compares the reward accumulation of three different RL solutions over 3000 episodes. DDPG and DQN are chosen as benchmarks as they are foundational in nature and widely used in literature for optimization of RIS coefficients and UAV trajectory. DA-DDPG collects 24\% and 97\% higher rewards as compared to DDPG and DQN for a learning rate of 0.0001. The standard DDPG, using a single actor network for both action types, performs notably better than DQN but is less efficient than DA-DDPG. The DDPG has a similar architecture to DA-DDPG but employs a single actor network for both discrete and continuous (thresholded) actions. While the simpler architecture can make it easier to train, DDPG struggles to handle the complexity of hybrid action spaces. DQN is limited to discrete outputs and shows the lowest performance among the three. These improvements highlight the significant advantage of DA-DDPG's separated actor network approach in handling hybrid action spaces, effectively balancing both continuous and discrete action learning. \ref{reward_vs_architecture} also illustrates the impact of the learning rate, with lr=0.01 being a bigger step size, converges to a lower reward for our complex problem. 
\vspace{-3mm}

\begin{figure}[b!]
    \centering
  \includegraphics[scale=0.45, trim=0cm 0cm 0cm 0.78cm, clip]{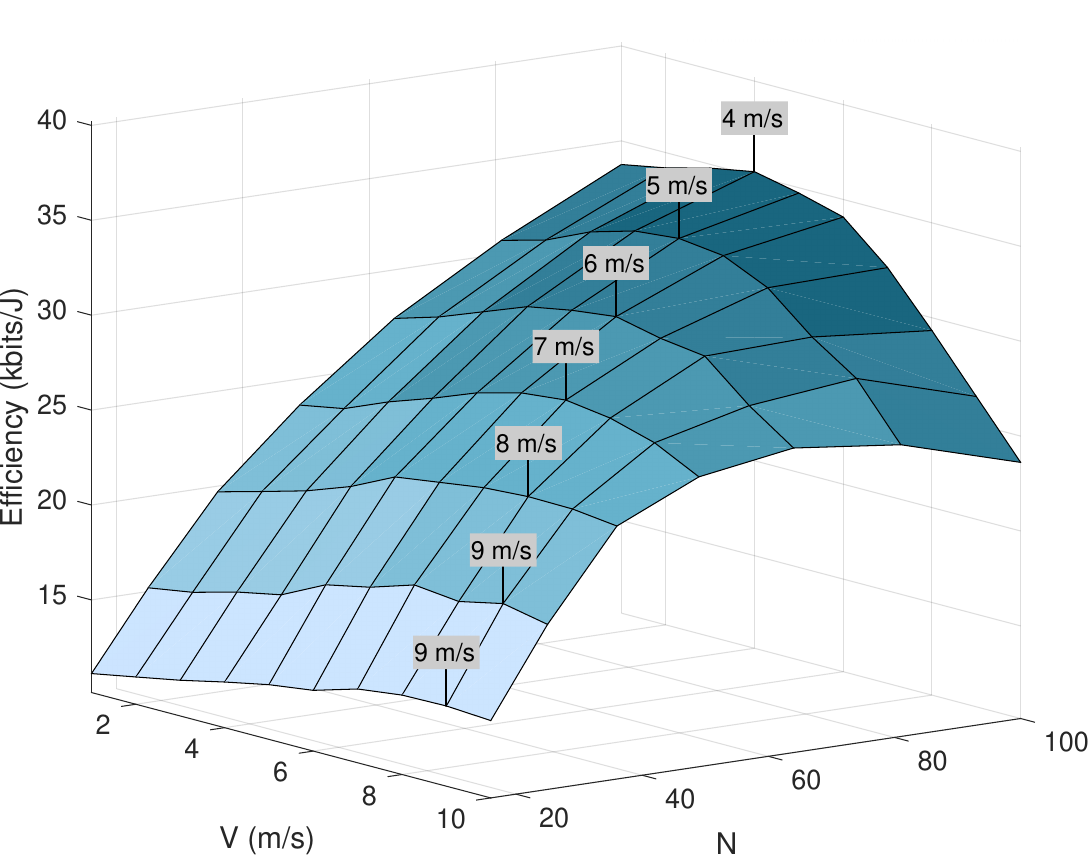}
  \caption{Efficiency variation with Aerial-STAR velocity and number of elements}
  \label{efficiency_plot}
\end{figure}

\subsubsection{STAR-RIS dimensions and UAV speed trade-off}Fig. \ref{efficiency_plot} sheds light on a complex relationship between $N$, Aeriel-STAR $V_{max}$ in 3D trajectory and the communication efficiency. The figure demonstrates that efficiency does not monotonically increase with $N$. For the same velocity, increasing $N$ increases the efficiency to a maximum before declining. This is because, although throughpu t increases with the increasing number of RIS elements, the RIS size and consequently the drag experienced by Aerial-STAR system also increases (quadratically with antenna dimensions \eqref{18}), thereby increasing the power consumption and decreasing communication efficiency. The plot also shows that as $V_{max}$ increases, the optimal velocity (corresponding to maximum efficiency) occurs at a lower $N$ value. This analysis demonstrates a trade-off between $N$, $V_{max}$, and overall communication efficiency, which must be carefully considered by design.\vspace{-5mm}

\subsubsection{Comparison of deployments strategies}
\begin{figure}[b]
    \centering
  \includegraphics[scale=0.55]{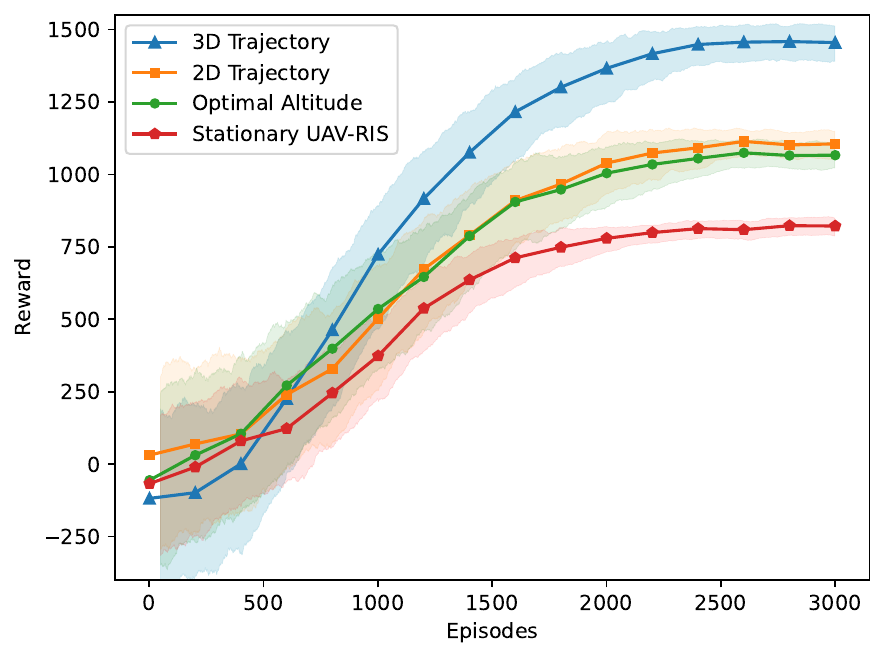}
  \caption{Reward for different Aeriel-STAR trajectories.}
  \label{reward_vs_trajectory}
\end{figure}

\begin{figure}
    \centering
    \begin{minipage}[b]{0.49\columnwidth}
        \centering
        \includegraphics[width=\textwidth]{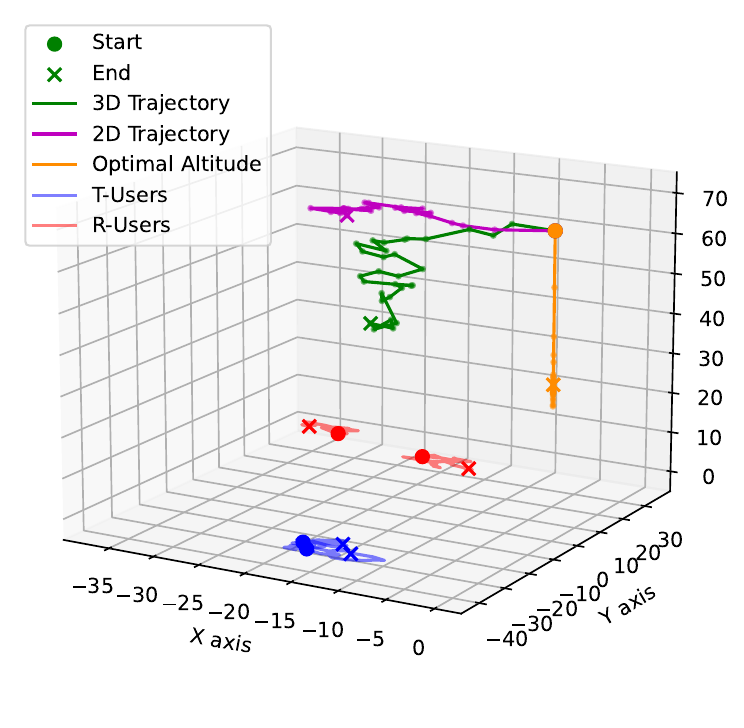}
    \end{minipage}
    \begin{minipage}[b]{0.49\columnwidth}
        \centering
        \includegraphics[width=\textwidth]{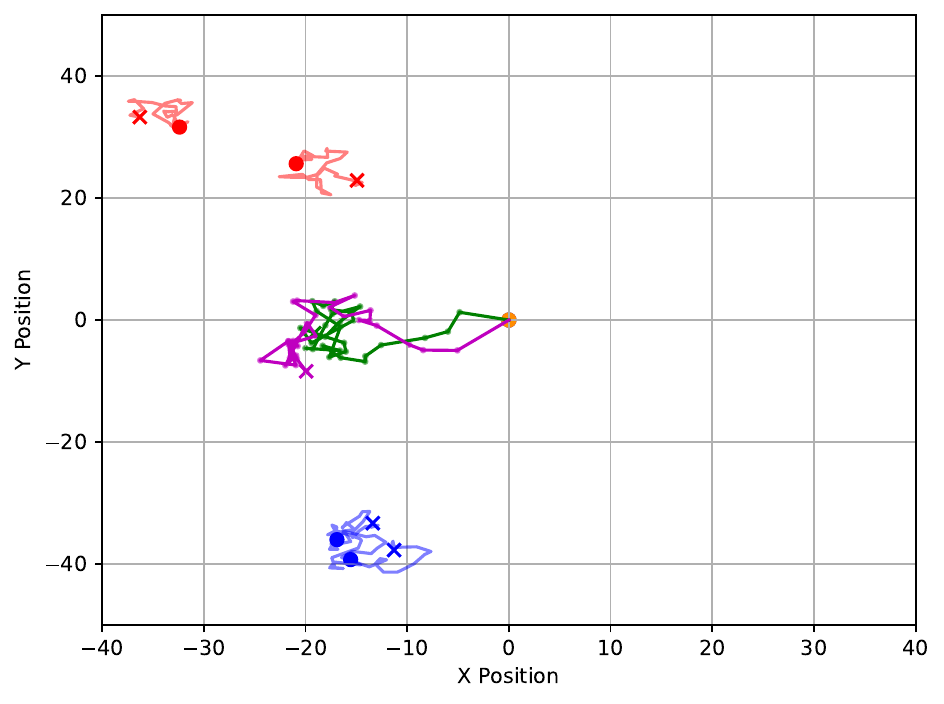}
    \end{minipage}
    \caption*{(a) Scenario 1: 3D (right) 2D (left) trajectories}
    
    
    \begin{minipage}[b]{0.45\columnwidth}
        \centering
        \includegraphics[width=\textwidth]{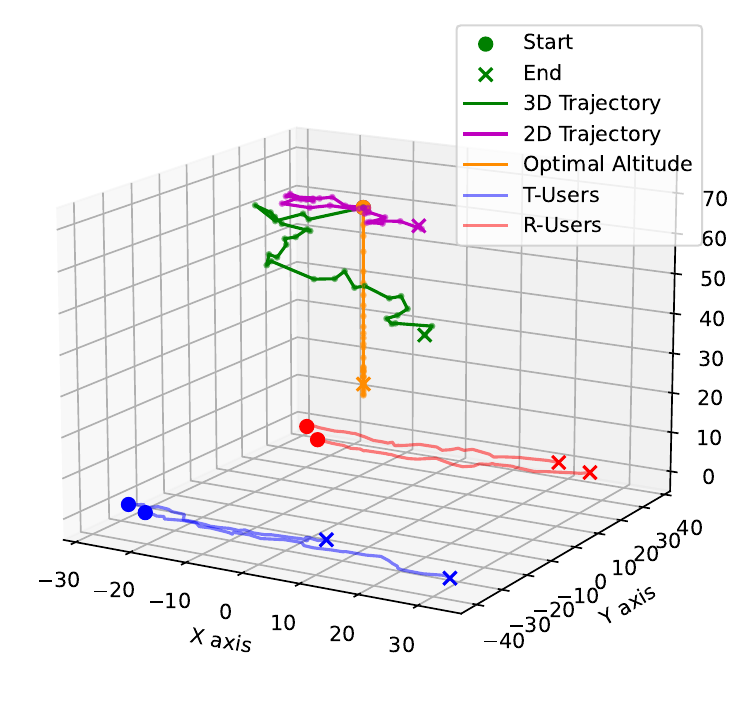}
    \end{minipage}
    \begin{minipage}[b]{0.45\columnwidth}
        \centering
        \includegraphics[width=\textwidth]{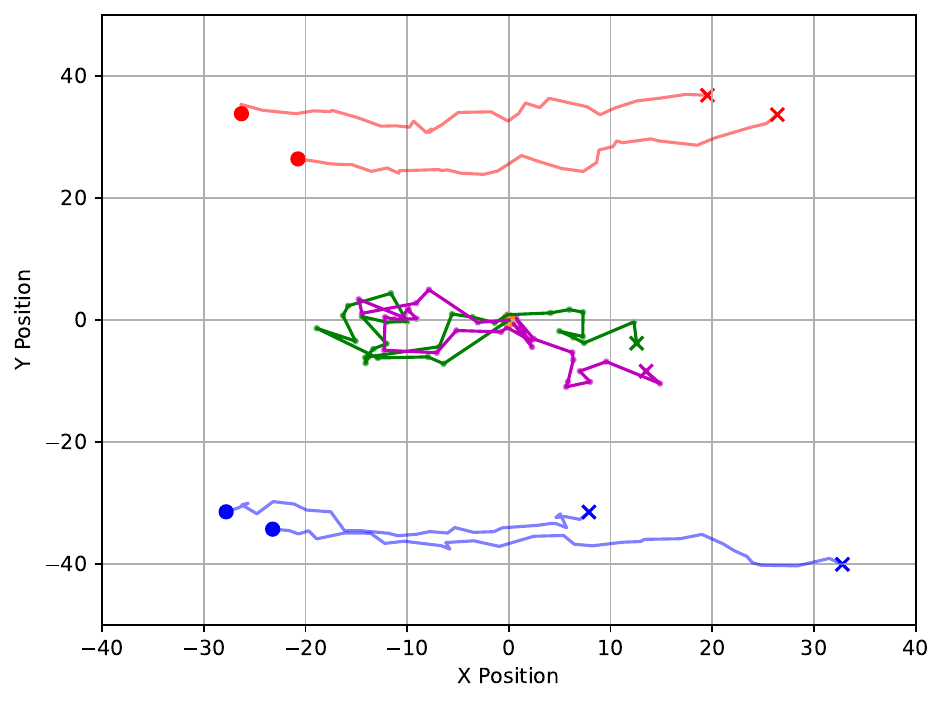}
    \end{minipage}
    \caption*{(b) Scenario 2: 3D (right) 2D (left) trajectories}
    \caption{Visualization of Aerial-Star trajectories in 3 different user movement scenarios - a) random movement and b) highly directional movement. It can be observed that 3D optimization allows the Aerial-STAR to closely follow the users' general direction of travel in 2D and optimize altitude while maintaining optimal coverage for both $\mathcal{T}$ and $\mathcal{R}$ user groups.}
    \label{fig:grid}
    {\vspace{-5mm}}
\end{figure}

Fig. \ref{reward_vs_trajectory} depicts the reward convergence for various Aerial-STAR deployment strategies, whereas Fig. \ref{fig:grid} helps visualizing all trajectories as per user movement. The 3D Trajectory strategy consistently achieves the highest reward (approximately 28\% improvement against the next best) by optimizing both 2D position and altitude, thus maximizing throughput and fairness. The Optimal 2D trajectory strategy shows lower performance, likely due to the added antenna drag without the benefit of altitude optimization to increase throughput. The Optimal Altitude strategy, while able to adjust the height, cannot optimize 2D position, resulting in sub-optimal performance due to the added power for vertical movement without the compensatory benefits of horizontal repositioning to optimize throughput. The Stationary deployment initially performs well due to lower power consumption, but ultimately converges with the lowest reward level. This highlights the trade-off between energy conservation and the ability to adapt to changing conditions and optimize throughput through repositioning. These results underscore the significant advantages of 3D positioning in Aerial-STAR systems, where each degree of freedom contributes to the overall system efficiency.
\vspace{-4mm}

\begin{figure}[b]
    \centering
  \includegraphics[scale=0.45]{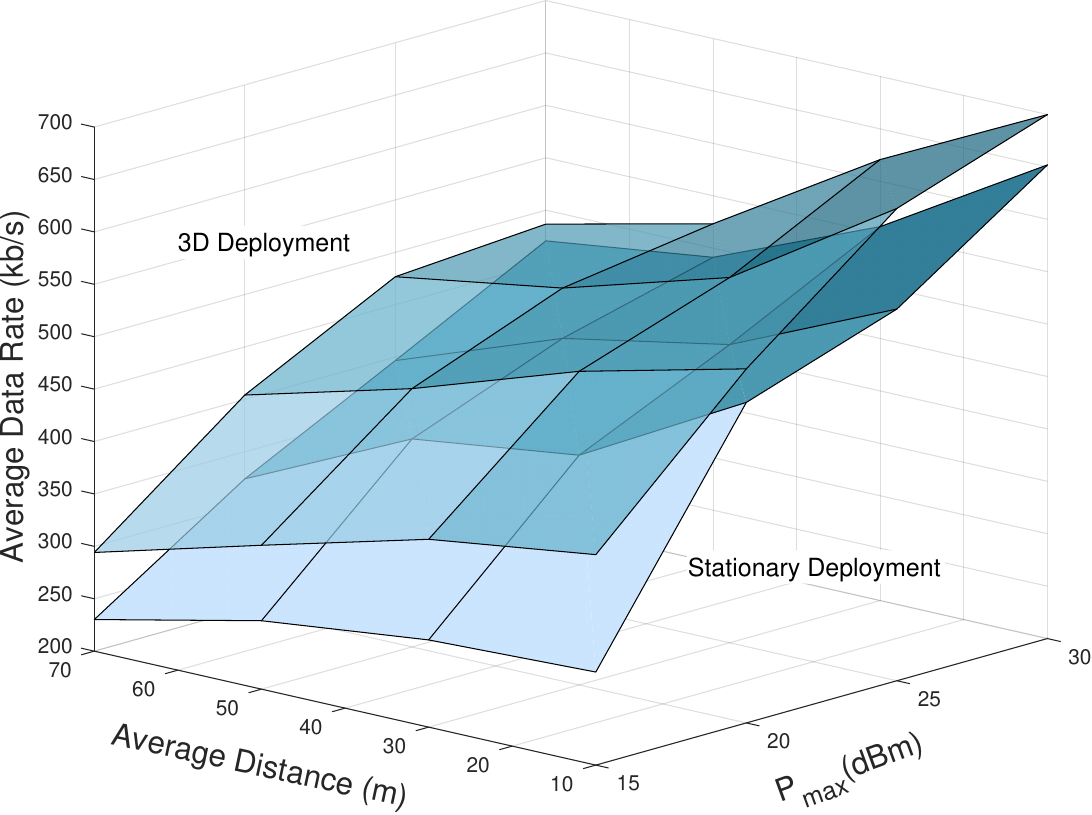}
  \caption{Average data rates per user for various $P_{max}$ and average distance between Aerial-STAR and UAV in the horizontal plane.}
  \label{Datarate_vs_Pmax_vs_dist}
\end{figure}

\subsubsection{Effect due to Transmission power and user distances} Fig. \ref{Datarate_vs_Pmax_vs_dist} reveals the difference in communication performance of a STAR-RIS deployed on UAV and its stationary counterpart against changing $P_{max}$ and average distance between STAR-RIS and users. Although the superior reward accumulation of 3D Deployment has been discussed in Fig. \ref{reward_vs_trajectory}, it is important to note that the reward function incorporates UAV power. In fact, a stationary STAR-RIS is likely to be installed on a fixed architecture so de-linking it from UAV power is necessitated. Our analysis emphasizes how much of that improvement in reward can be attributed to superior communication performance. Across various scenarios of user distances and maximum transmit powers at the base station, 3D deployment consistently outperforms stationary deployment due to its ability to optimize position in three-dimensional space. The 3D deployment maintains its superiority for increasing values of $P_{max}$ and average distance. For the lowest distance and power combination, 3D deployment outperforms stationary deployment by 38\%. It should be noted that this performance gap narrows at higher powers and greater distances - reducing to only 4\% for the highest distance and power combination. In the latter scenario, the impact of small position adjustments in Aerial-STAR becomes less pronounced relative to the larger user-UAV distances, with height optimization becoming the primary differentiator. Additionally, as transmit power increases, the extended range of stationary deployed RIS further minimizes the performance difference between the two approaches. It can be concluded that Aerial-STAR deployment could be favored over stationary deployments especially when the users in $\mathcal{R}$ and $\mathcal{T}$ regions are not very distant.\vspace{-4mm}

\begin{figure}[b]
    \centering
  \includegraphics[scale=0.55]{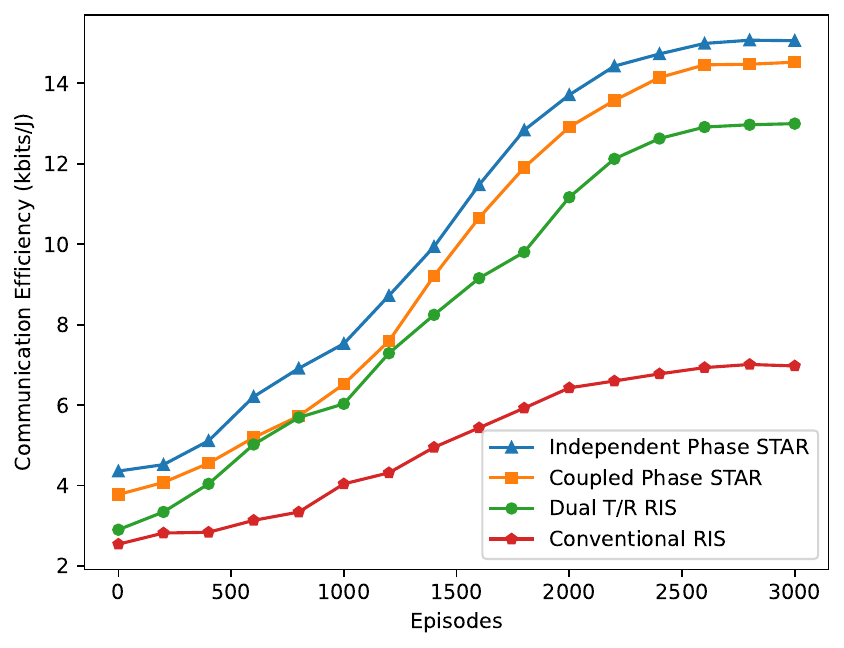}
  \caption{Communication efficiency plot while training for STAR-RIS, Dual T/R RIS and conventional Reflecting RIS}
  \label{RIS_type_vs_eff}

\end{figure}

\subsubsection{Aerial-STAR performance with various STAR-RIS types}Comparison of various RIS schemes is shown in Fig. \ref{RIS_type_vs_eff}. The Independent Phase STAR RIS, as the ideal scenario, outperforms the Coupled Phase STAR RIS by a small margin due to the practical limitations of phase control in the latter. However, the coupled phase STAR RIS shows a 22\% improvement over the Dual T/R RIS, thanks to its ability to simultaneously serve both $\mathcal{T}$ and $\mathcal{R}$ users. In contrast, the Dual T/R RIS is configured as 2 RIS, each serving $\mathcal{R}$ and $\mathcal{T}$ regions, with a total number of elements the same as Aerial-STAR for a fair comparison. The Dual T/R RIS functions as a `Mode Switching' RIS with a fixed T/R element ratio and lacks the flexibility of the STAR-RIS, making the STAR-RIS more effective in dynamic environments. The conventional reflecting-only RIS supports only $\mathcal{R}$ users, failing to assist $\mathcal{T}$ users on the opposite side and resulting in minimum efficiency.

\subsubsection{Effect of adding HFI to reward function}
\begin{figure}
    \centering
  \includegraphics[scale=0.55]{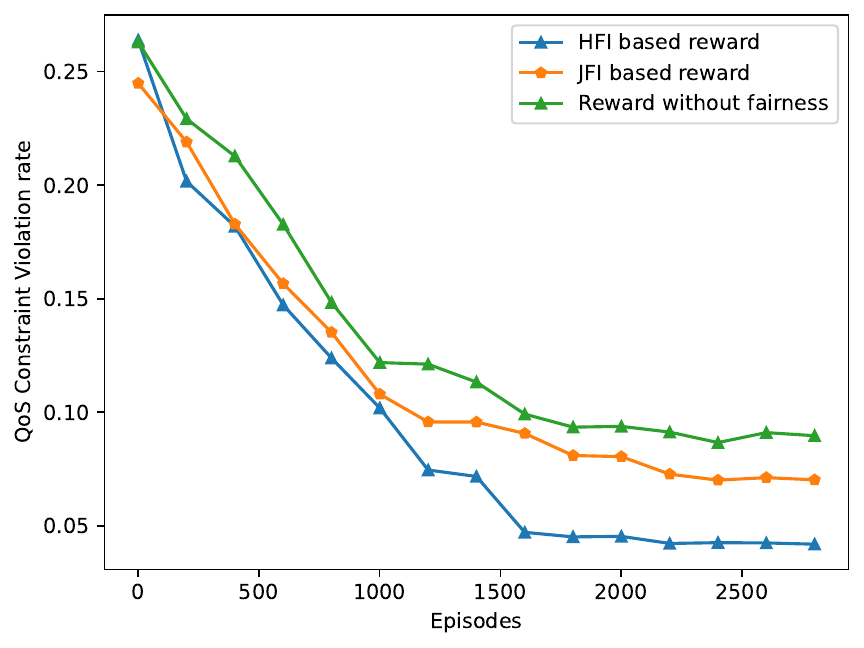}
  \caption{Minimum data rate QOS violation rate graph for comparing HFI and JFI.}
  \label{QOS_vs_index_function}
\end{figure}

The impact of incorporating fairness indices into the reward function is shown in Fig.~\ref{QOS_vs_index_function}. It compares the minimum data rate QoS constraint violation rates for different reward formulations. The graph reveals that both the HFI and JFI-based rewards significantly outperform the baseline reward without fairness considerations. Notably, the HFI-based reward consistently achieves the lowest QoS constraint violation rate, rapidly decreasing to below 0.10 within the first 500 episodes and exhibiting a 41\% final improvement as compared to the JFI-based reward function. This is because the HFI imposes a higher penalty in scenarios with high user data rate disparity as compared to JFI, which shows a similar trend but maintains a slightly higher violation rate throughout. These results underscore the efficacy of incorporating fairness metrics in the reward function. We demostrate that HFIis capable of serving more users with minimum data rate QoS constraint, leading to more consistent network performance across all users.

\section{Conclusion}
This study analyzes the effect of deploying a coupled phase shift modeled STAR RIS on a UAV for downlink communications. Optimizing the UAV trajectory, base station beamforming vectors and coupled TRCs was a challenging problem as it required discrete and continuous-valued outputs. For this, we propose a DA-DDPG algorithm with two actor networks (one each for discrete and continuous actions) to optimize the communication efficiency of the Aerial-STAR system. We also propose a new Harmonic Fairness Index and show that incorporating it in our reward function improves the fairness between users. The results show that 1) DA-DDPG outperforms the DDPG and DQN algorithms by 24\% and 97\%, respectively, 2) 3D deployment results in the highest reward other deployment strategies, 3) coupled phase STAR performs better than Dual T/R RIS and conventional RIS and 4) increasing the number of elements in a UAV deployed RIS would not necessarily increase the communication efficiency.

\begin{figure}
    \centering
  \includegraphics[scale=0.35]{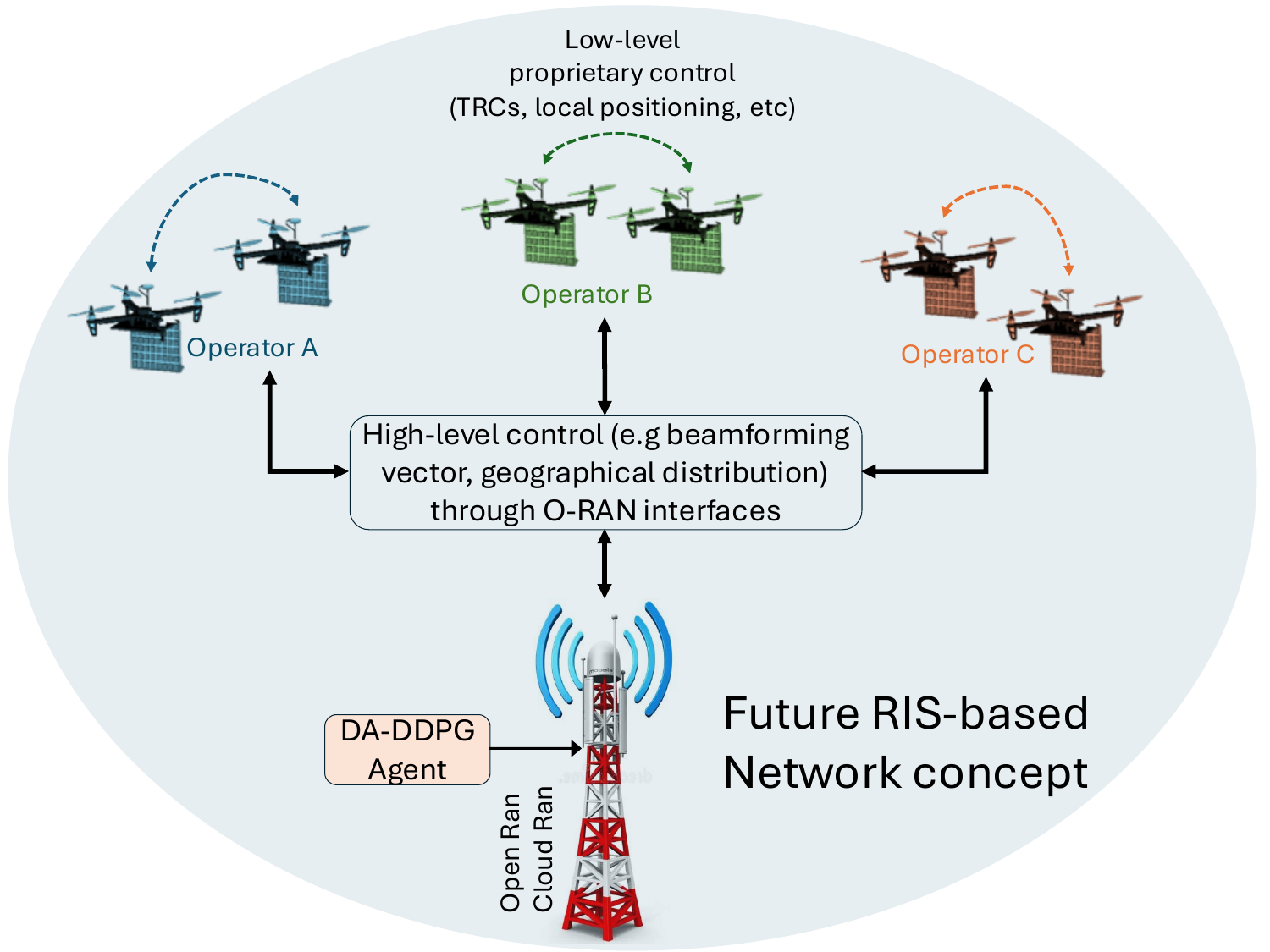}
  \caption{A model for Open Ran operation to facilitate multi-vendor Aerial-RIS enabled networks.}
  \label{future_network_ris}
\end{figure}

In future networks (envisioned in Fig. \ref{future_network_ris}), multiple third-party Aerial-STARs could simultaneously provide services, with base stations operated independently. This scenario necessitates a robust handshaking mechanism between services, aligning with the Open RAN concept highlighted in the European Union's 6G research initiative, Hexa-X \cite{hexa-x}, and Next G Alliance's ``Roadmap to 6G" \cite{nextg6g} report as a key enabler for future network architectures. In this setup, an expanded version of the DA-DDPG agent could be integrated into an Open RAN-compatible base station architecture, such as Ericsson's Cloud RAN solution, which is expected to support the O-RAN fronthaul interface~\cite{ericsson_openran_forward}. The base station could provide high-level control by managing broader Aerial-STAR positioning to maximize coverage through standardized Open RAN interfaces, while low-level control of each cluster of UAVs could be managed locally by proprietary third-party softwares. These principles are also reflected in 3GPP's technical report TR 38.801 \cite{3gpp_38801}, which outlines architectures supporting functional splits in radio access networks. This hierarchical approach enables flexible and robust network optimization within an open, multi-vendor ecosystem, allowing for efficient coordination of multiple Aerial-STAR systems.

\balance
{\footnotesize
\bibliographystyle{unsrt}
\bibliography{references}}

\end{document}